\documentclass{article}

\usepackage{PRIMEarxiv}

\usepackage{abstract}
\usepackage[utf8]{inputenc} 
\usepackage[T1]{fontenc}   
\usepackage{hyperref}       
\usepackage{url}            
\usepackage{booktabs}       
\usepackage{amsfonts}       
\usepackage{amsmath}        
\usepackage{amsthm}         
\usepackage{nicefrac}       
\usepackage{microtype}      
\usepackage{lipsum}
\usepackage{fancyhdr}       
\usepackage{graphicx}       
\usepackage{subcaption}
\usepackage{braket}
\graphicspath{{media/}}     

\newtheorem{proposition}{Proposition}[section]

\newtheorem{remark}{Remark}[section]

\title{Solving Systems of Linear Equations: HHL from a Tensor Networks Perspective
 }

\author{
  Alejandro Mata Ali \\
  Instituto Tecnológico de Castilla y León, Burgos, Spain\\
  \texttt{alejandro.mata.ali@gmail.com} \\
  \And
  Iñigo Perez Delgado \\
  i3B Ibermatica, Parque Tecnológico de Bizkaia \\
  Ibaizabal Bidea, Edif. 501-A \\
  48160 Derio, Spain\\
  \texttt{iperezde@ayesa.com} \\
  \And
  Marina Ristol Roura \\
  i3B Ibermatica, Parque Tecnológico de Bizkaia \\
  Ibaizabal Bidea, Edif. 501-A \\
  48160 Derio, Spain\\
  \texttt{mristol@ayesa.com} \\
  \And
  Aitor Moreno Fdez. de Leceta \\
  Quantum Technologies and Systems Unit,\\
  LKS Next, MONDRAGON Corporation, Goiru 7,\\
  20500 Arrasate-Mondragón, Gipuzkoa, Spain\\
  \texttt{aitormoreno@lksnext.com} \\
  \And
  Sebasti\'an V. Romero \\
  TECNALIA, Basque Research and Technology Alliance (BRTA), 48160 Derio, Spain\\
  \texttt{sebastian.vidal@tecnalia.com} \\
}

\begin{document}
\maketitle

\begin{abstract}
This work presents a new approach for simulating the HHL linear systems of equations solver algorithm with tensor networks. First, a novel HHL in the qudits formalism, the generalization of qubits, is developed, and then its operations are transformed into an equivalent classical HHL, taking advantage of the non-unitary operations that they can apply. The main novelty of this proposal is to perform a classical simulation of the HHL as efficiently as possible to benchmark the algorithm steps according to its input parameters and the input matrix. The algorithm is applied to three classical simple simulation problems, comparing it with an exact inversion algorithm, and its performance is compared against an implementation of the original HHL simulated in the Qiskit framework, providing both codes. It is also applied to study the sensitivity of the HHL algorithm with respect to its hyperparameter values, reporting the existence of saturation points and maximal performance values. The results show that this approach can achieve a promising performance in computational efficiency to simulate the HHL process without quantum noise, providing a higher bound for its performance.
\end{abstract}

\tableofcontents
\section{Introduction}
The solution of linear equation systems $A\vec{x}=\vec{b}$ is a fundamental problem in many areas of science and engineering. Classical methods for solving these equations, such as \textit{Gaussian elimination} and \textit{LU decomposition} \cite{LU}, have been widely used and optimized for decades~\cite{Random_LU,LargeLU}. However, as the size of the system grows, classical methods become computationally expensive and inefficient. One of the most efficient classical methods in the sparse Hermitian positive definite setting is the \textit{conjugate gradient method} (CG) \cite{CG,CGGPU}, which has a complexity of $\mathcal{O}\left(Ns\sqrt{\kappa}\log\left(\frac{1}{\epsilon}\right)\right)$ for a matrix $N\times N$ with a maximum of $s$ non-zero elements per row, $\kappa \equiv \frac{\lambda_{max}}{\lambda_{min}}$, $\lambda_{min},\lambda_{max}>0$ being the extreme eigenvalues of $A$ and $\epsilon$ the error.

Quantum computers offer the potential to solve some challenging problems more efficiently than classical computers. In particular, the \textit{HHL algorithm} proposed by Harrow, Hassidim, and Lloyd in 2008 \cite{HHL,HHL_step} is a method for solving linear equations that runs in polynomial time, where the polynomial depends logarithmically on the size of the system. It is intended for the calculation of the expectation values in $\mathcal{O}\left(\log(N)s^2\kappa^2/\epsilon^3\right)$, as it loses its advantage in the case of extracting the explicit solution, which is produced in time $\mathcal{O}\left(\log(N)s^2\kappa^2/\epsilon\right)$. This algorithm has three main hyperparameters: $\tau$ the evolution time, which rescales the eigenvalues of the problem, $n_c$ which determines the precision in the eigenvalues, and $C$ which rescales the eigenvalue inversion step. The dependence of the algorithm performance with respect to these three hyperparameters, even in the noise-free regime, is an important open question.

However, the current Noisy intermediate-scale quantum (NISQ) state of the quantum hardware limits the testing of this algorithm, restricting its benchmarking to classical simulators. For the HHL algorithm, the most feasible simulation framework is the statevector simulation. This framework has several limitations, such as the exponential increase in the computational resources required with respect to the number of qubits, which is approximately $\mathcal{O}\left(N^2\left(\frac{\kappa}{\epsilon}\right)\log\left(\frac{\kappa}{\epsilon}\right) + N\left(\frac{\kappa}{\epsilon}\right)^2\right)$ in the best possible case and $\mathcal{O}\left(N^2\left(\frac{\kappa}{\epsilon}\right)\log\left(\frac{\kappa}{\epsilon}\right) + N\left(\frac{\kappa}{\epsilon}\right)^3\right)$ in the most common situation (as proved in Sec.~\ref{sec: background}). This is due to the simulation of a $n+n_c$ qubit circuit with a general Hamiltonian evolution and the $n_c$ steps of the quantum phase estimation, $n=\lceil\log_2 N\rceil$ and $n_c=O(\log_2(\kappa/\epsilon))$ being the number of clock qubits required for the precision of the eigenvalues.

Recently, there has been growing interest in using qudits \cite{Qudits} and tensor networks \cite{Tensor_Network,ORUS2014117} to implement different quantum algorithms. Qudits are generalized qubits with more than 2 basis states. Tensor networks are classical representations of tensor algebra equations, providing an efficient way to represent and manipulate certain types of high-dimensional systems, such as quantum states with low entanglement \cite{TNQML}, or compressing machine learning models \cite{TNML,MLComp}, enabling quick computations with classical computers \cite{Multigrid,Navier,Survey_TN_Appl}. The tensor networks framework is also applied to circuit simulation~\cite{TN_for_QC}, applying tensor compression, but it scales exponentially with respect to the entanglement. Moreover, if it is applied directly without other optimizations adapted specifically to the circuit simulated, the amount of resources can scale even more, making it even slower than the statevector simulation.

In this paper, a novel approach is proposed for simulating the HHL linear systems of equations solver algorithm using qudits and tensor networks, providing an efficient way to test the higher bound in the possible performance of the algorithm. This algorithm runs in $\mathcal{O}(N\kappa^2/\epsilon^2+N^2\kappa/\epsilon+\kappa^3/\epsilon^3)$. It is demonstrated how this approach can be used to simulate the HHL performance with a large number of variables, and the runtime of this approach is compared with existing quantum and classical methods for both solving linear systems of equations and simulating the HHL algorithm classically. The main contributions of this work are the following.
\begin{itemize}
    \item The formulation of a qudit HHL algorithm.
    \item The formulation and implementation of a classical tensor network algorithm to classically simulate the best expected performance of the HHL algorithm.
    \item Comparison of this algorithm against well-known classical algorithms in different scenarios.
    \item A numerical analysis of the sensitivity of the HHL algorithm with respect to its hyperparameters.
\end{itemize}

These results show that this approach can achieve a promising performance in computational efficiency to simulate the HHL process without quantum noise. However, it does not improve the performance of the state-of-the-art best-known linear solver algorithms, and it is not intended to do so.

The paper is organized as follows. Initially, Sec.~\ref{sec: background} presents a concise overview of classical algorithms designed to solve linear equations, as well as the HHL algorithm, omitting detailed quantum computing concepts, and an innovative qudit algorithm in qudits that enhances the original HHL is introduced. Following this, Sec.~\ref{sec: tensor network} describes the classical tensor network approach to simulate the optimal HHL performance and analyzes its complexity. Sec.~\ref{sec: comparison} is devoted to comparing the novel algorithm with the conjugate gradient method, the traditional HHL, and the statevector simulation of the HHL. In the final section, in Sec.~\ref{sec: simulation} the algorithm is applied to three simulation problems, contrasting it with an exact inversion algorithm, its performance is evaluated relative to an implementation of the HHL algorithm simulated via the Qiskit framework, and its sensitivity of the HHL performance respect to its hyperparameters. 
All code is available in the GitHub repository \href{https://github.com/DOKOS-TAYOS/Tensor_Networks_HHL_algorithm.git}{https://github.com/DOKOS-TAYOS/Tensor\_Networks\_HHL\_algorithm.git}, it can be openly tested online on the Streamlit website \href{https://tensornetworks-hhl-algorithm.streamlit.app/}{https://tensornetworks-hhl-algorithm.streamlit.app/}, and there is a brief spanish explanation of the paper in the video \href{https://www.youtube.com/watch?v=bxmZDIblxmI&list=PLL6OBevU4q9CfaXXw4aivzPFhNxmG6NpW&index=1}{
Algoritmo HHL con Tensor Networks y qudits}.

\section{Background}\label{sec: background}
There exist several algorithms to solve systems of linear equations, but only a few interesting ones will be introduced. All of them solve the system of linear equations
\begin{equation}
    A\vec{x} = \vec{b},
\end{equation}
where $A$ is an invertible matrix $N \times N$, $\vec{x}$ is the vector to be obtained, and $\vec{b}$ is another vector, both of dimension $N$.

The first is \textit{Gaussian elimination}, which consists of reducing the augmented matrix $[A|b]$ to a matrix of the row echelon through row addition operations and transforming it into a diagonal matrix with the required solutions. Its computational complexity is $\mathcal{O}(N^3)$. It is widely used in small cases, but when matrices become too large, other methods are applied because of its prohibitive cubic cost~\cite{Numerical_LA}. Another extended algorithm is the \textit{LU decomposition}~\cite{LU}, which can be viewed as another way to perform Gaussian elimination. It decomposes the matrix $A$ into two matrices $L$ (lower triangular) and $U$ (upper triangular) that satisfy $A=LU$. It has the same complexity as Gaussian elimination. Following the same line, there is the \textit{Cholesky decomposition}~\cite{cholesky}, which consists of decomposing a Hermitian positive definite matrix $A$ into a lower triangular matrix $L$ with real positive diagonal entries that satisfy $A=LL^*$. Its complexity is $\mathcal{O}(N^3)$, but it has half the cost of LU decomposition. This method is widely used for Monte Carlo simulation, but not in general inversion due to the Hermitian positive definite matrix restriction.

In the field of iterative methods, the most simple one is the \textit{Jacobi algorithm}~\cite{iterative_matrix_equations}. First, the matrix $A$ is decomposed into three matrices $D$ diagonal, $L$ lower triangular, and $U$ upper triangular such that $A=D+L+U$. For the $k$-th iteration, the transformation
\begin{equation}
    \vec{x}^{(k+1)} = D^{-1} \left( \vec{b} - (L + U) \vec{x}^{(k)} \right)
\end{equation}
is applied until it converges. Its complexity is $\mathcal{O}(N^2)$ for each step in the dense case and can take advantage of the sparsity of $A$, but the number of necessary steps is not well-known. For this reason, it is used in large problems.

An improvement of this method is the \textit{Gauss-Seidel} algorithm for positive definite matrices, or strictly diagonally dominant matrices, with no zero diagonal elements. In this case, the iteration is made with
\begin{equation}
    x_i^{(k+1)} = \frac{1}{a_{ii}} \left( b_i - \sum_{j=1}^{i-1} a_{ij} x_j^{(k+1)} - \sum_{j=i+1}^{n} a_{ij} x_j^{(k)} \right).
\end{equation}
This method has the advantage in the storage, allowing one to rewrite the first iteration vector, which enables to address larger problems, but it is much harder to implement in parallel.

Another iterative algorithm is the well-known \textit{conjugate gradient method}~\cite{CG} (CG) for large sparse positive definite matrices. Two non-zero vectors $\vec{v}$ and $\vec{w}$ are conjugate with respect to $A$ if
\begin{equation}
    \vec{v}^T A \vec{w} = 0.
\end{equation}
If a set $P=\lbrace \vec{p}_0, \vec{p}_1, \dots ,\vec{p}_{N-1}\rbrace$ of mutually conjugate vectors with respect to $A$ is defined, then $P$ forms a basis on $\mathbb{R}^N$ and
\begin{equation}
    \vec{x} = \sum_{i=0}^{N-1} \alpha_i \vec{p}_i \implies A \vec{x} = \sum_{i=0}^{N-1} \alpha_i A \vec{p}_i.
\end{equation}

Multiplying the problem by the left with $\vec{p}_j^T$
\begin{equation}
    \vec{p}_j^T\cdot \vec{b} = \vec{p}_j^T A \vec{x} = 
    \sum_{i=0}^{N-1} \alpha_i \vec{p}_j^T A \vec{p}_i = \alpha_j \vec{p}_j^T A \vec{p}_j,
\end{equation}
then
\begin{equation}
    \alpha_j = \frac{\vec{p}_j^T\cdot \vec{b}}{\vec{p}_j^T A \vec{p}_j}.
\end{equation}
So, a set $P$ is needed, and then the $\alpha_j$ values are computed. However, if good $\vec{p}_i$ vectors are chosen, it may not be necessary to compute all of them to solve the problem. So, the problem will be solved approximately. The algorithm starts with an initial guess $\vec{x}_0$, and has a function to minimize
\begin{equation}
    f(\vec{x}) = \frac{1}{2} \vec{x}^T A \vec{x} - \vec{x}^T\cdot \vec{b}.
\end{equation}
Due to $\nabla f(\vec{x}) = A \vec{x} - \vec{b}$, it takes $\vec{p}_0= \vec{b} - A \vec{x}_0$. $\vec{r}_j$ being the residual of the $j$-th step, and the negative gradient of $f$ in $\vec{x}_j$
\begin{equation}
    \vec{r}_j = \vec{b} - A \vec{x}_j.
\end{equation}

To obtain a vector $\vec{p}_j$ conjugate to the other ones, the transformation
\begin{equation}
\vec{p}_j = \vec{r}_j - \sum_{i < j} \frac{\vec{r}_j^T A \vec{p}_i}{\vec{p}_i^T A \vec{p}_i} \vec{p}_i,
\end{equation}
is applied, and $\vec{x}_{j+1}=\vec{x}_j+\alpha_j\vec{p}_j$ is updated with
\begin{equation}
\alpha_j = \frac{\vec{p}_j^T\cdot \left( \vec{b} - A \vec{x}_j \right)}{\vec{p}_j^T A \vec{p}_j} = \frac{\vec{p}_j^T\cdot \vec{r}_j}{\vec{p}_j^T A \vec{p}_j}
\end{equation}
until the residual is small enough.

Similar methods such as the \textit{generalized minimal residual method}~\cite{GMRES} and the \textit{biconjugate gradient stabilized method}~\cite{BICGSTAB} are based on similar ideas as improvements for more general problems. All of these algorithms require matrix-vector multiplications, so their complexity is at least linear in the dimension of the problem. However, these faster state-of-the-art algorithms are only for specific problems, so they are presented only for extreme performance comparison against the following HHL algorithm. This is the reason because the common use libraries as NumPy or PyTorch make use of LU factorization for inversion problems to be able to solve general problems.

In this context, the \textit{quantum linear solver algorithm} HHL~\cite{HHL} provides a computational advantage. Now, the standard HHL algorithm in qubits is introduced in order to better understand the algorithm that will be formalized.

\begin{figure}[h]
    \centering
    \begin{subfigure}[b]{0.48\linewidth}
        \centering
        \includegraphics[width=\linewidth]{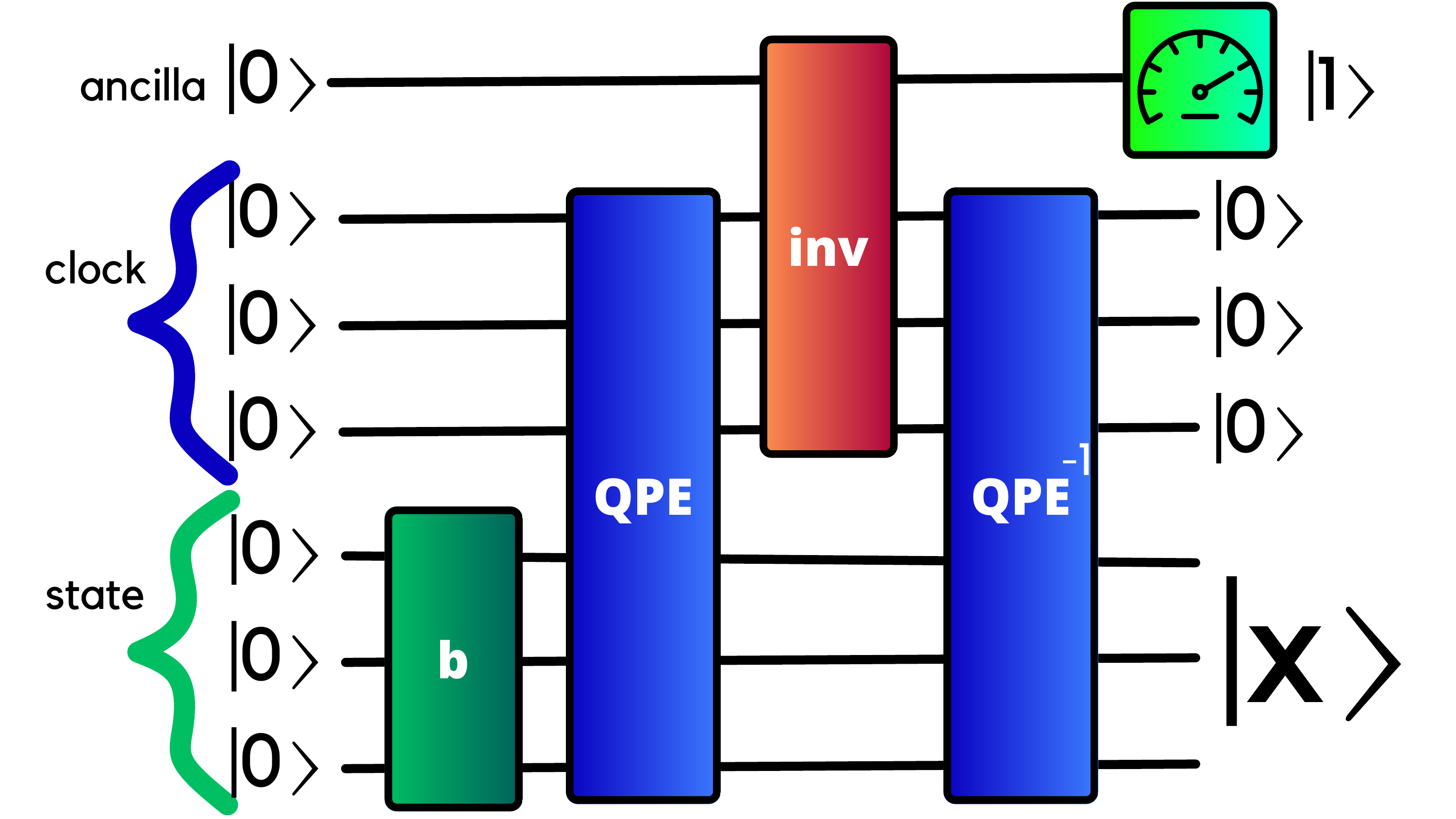}
        \caption{Quantum HHL in qubits with $n=n_c=3$.}
        \label{fig:HHL Quantum}
    \end{subfigure}
    \begin{subfigure}[b]{0.48\linewidth}
        \centering
        \includegraphics[width=\linewidth]{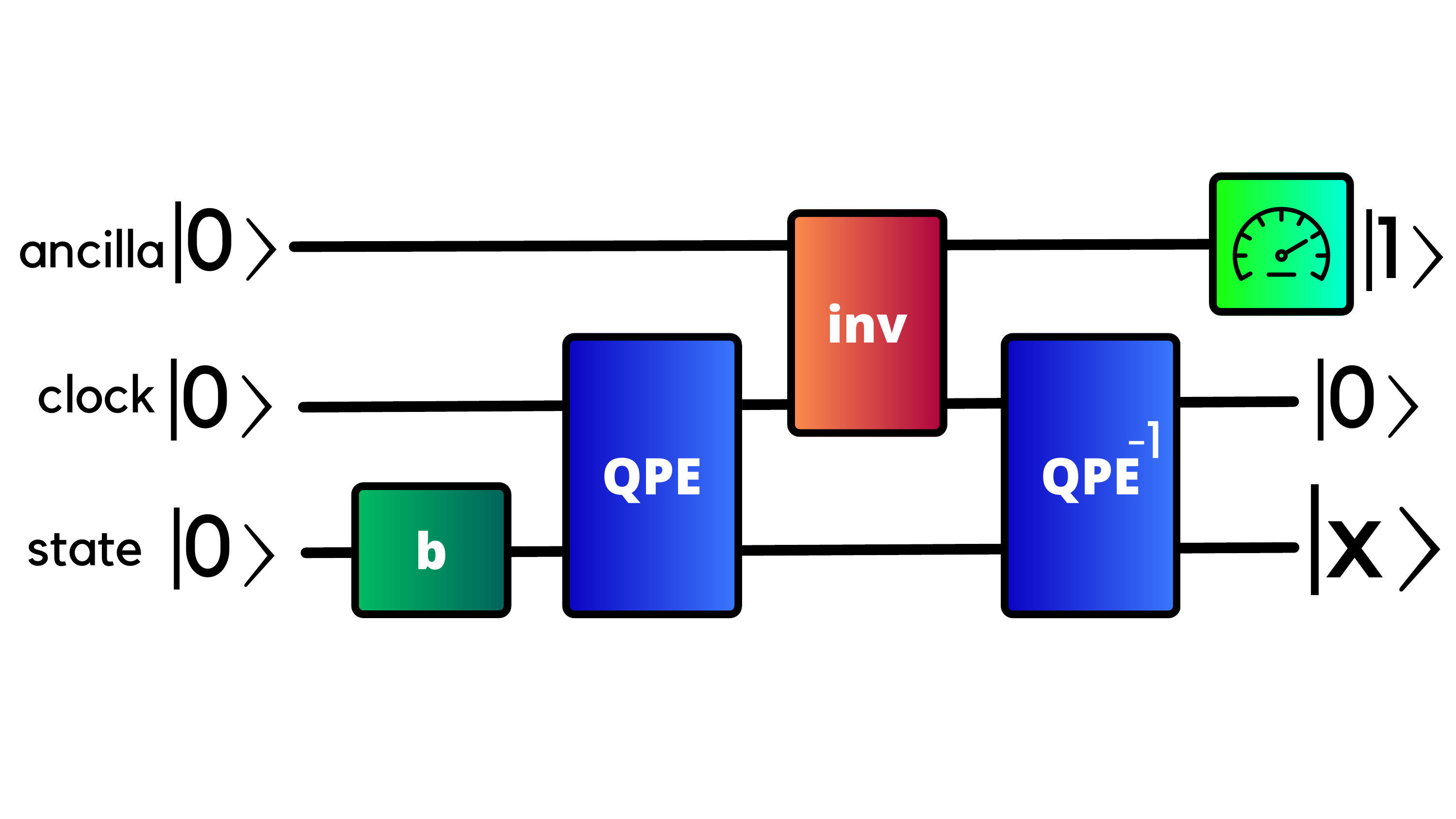}
        \caption{Quantum HHL in qudits.}
        \label{fig:HHL Qudits}
    \end{subfigure}
    \caption{Comparison of HHL algorithm implementations.}
    \label{fig:HHL comparison}
\end{figure}

For this algorithm, $n=\lceil\log_2 N\rceil$ qubits are needed to encode the vector $\vec{b}$, $n_c$ clock qubits to encode the possible eigenvalues of $A$ and one auxiliary qubit for the inversion with respect to these eigenvalues. The whole circuit can be summarized in Fig. \ref{fig:HHL Quantum}.

The state $\vec{b}$ is encoded in the amplitudes of the $n$ state qubits
\begin{equation}
    |b\rangle = \sum_{i=0}^{2^n-1} b_i |i\rangle = \sum_{i=0}^{2^n-1} \beta_i |u_i\rangle,
\end{equation}
where $b_i$ are the normalized components of the vector $\vec{b}$, $|i\rangle$ are the computational bases states and $|u_i\rangle$ the eigenvector associated with the eigenvalue $\lambda_i$ of $A$. An operator $b$ is applied to initialize it. It is important to note that the difference between $N$ and $2^n$ will be completed with zeros in the vector and a matrix proportional to the identity in $A$, wasting resources. Moreover, a method is needed to generate this state $|b\rangle$ or the $b$ operator~\cite{State_Preparation,state_prep}.

The second step is to calculate the evolution operator
\begin{equation}
    U_{\mu,\tau}=e^{2\pi i \tau A'/\mu},
\end{equation}
where $\tau$ is a hyperparameter that rescales the eigenvalues for the next step, $\mu=2^{n_c}$ is the number of phase bins encoded by the clock register, and the matrix $A'=A$ if $A$ is Hermitian, and
\begin{equation}
    A' = \left(
    \begin{matrix}
    0 & A \\
    A^\dagger & 0
    \end{matrix}
    \right)\label{eq: hermitian}
\end{equation}
if it is not. In this case, the problem is 
\begin{equation}
    A' \left(
    \begin{matrix}
    0 \\
    \vec{x}
    \end{matrix}
    \right)=
    \left(\begin{matrix}
    \vec{b} \\
    0
    \end{matrix}
    \right).\label{eq: non hermitian}
\end{equation}
If $A$ is invertible and has singular value decomposition $A=\sum_j \sigma_j \ket{v_j}\!\bra{w_j}$, then $A'$ is Hermitian with eigenvectors $\frac{1}{\sqrt{2}}(\ket{v_j},\ket{w_j})^T$ and $\frac{1}{\sqrt{2}}(\ket{v_j},-\ket{w_j})^T$, associated with eigenvalues $+\sigma_j$ and $-\sigma_j$, respectively. Therefore, the relevant condition number in the non-Hermitian reduction is
\begin{equation}
    \kappa(A')=\frac{\sigma_{max}(A)}{\sigma_{min}(A)}=\kappa_2(A),
\end{equation}
so the embedded problem depends on the singular values of $A$, not on its generally complex eigenvalues. It is assumed that $U_{\mu,\tau}$ can be calculated and implemented efficiently~\cite{Sparse_H}.

With this operator $U_{\mu,\tau}$ presented, a \textit{Quantum Phase Estimation} (QPE)~\cite{Nielsen} is performed to encode the eigenvalues of $A'$ in the clock qubits. Since there are $n_c$ qubits, $\mu=2^{n_c}$ possible clock values can be encoded. In this convention, the QPE output index $d$ approximates the rescaled eigenvalue $\tau\lambda$ modulo $\mu$, so the choice of $\tau$ and $\mu$ must be compatible with the spectral window of the problem. For this reason, $n_c$ (and therefore $\mu$) is a hyperparameter for resolution, since it has to be chosen based on the properties of the problem, usually the condition number of the matrix. Typically, under the usual HHL spectral-separation assumptions, $n_c=O(\log_2(\kappa/\epsilon))$, then $\mu=O(\kappa/\epsilon)$~\cite{HHL}.

Now, an inversion operator is applied, which rotates the probability of the auxiliary qubit so that it is divided by the value of the eigenvalue encoded by the QPE.

The next step is to make a post-selection, keeping only the state if the auxiliary qubit outputs a $\ket{1}$, followed by an inverse QPE to clean the eigenvalue qubits.

In the end, the result is the $\vec{x}$ state normalized in the amplitudes, 
\begin{equation}
    |x\rangle = \frac{1}{\mathcal{N}}\sum_{i=0}^{2^n-1} \frac{\beta_i}{\lambda_i} |u_i\rangle = \frac{1}{\mathcal{N}}\sum_{i=0}^{2^n-1} x_i |i\rangle ,
\end{equation}
with a normalization constant $\mathcal{N}$ and omitting the ancilla and clock qubits.

To obtain the full state vector, measuring only in the computational basis gives access to the probabilities $|x_i|^2$, not to the signs or relative phases of the amplitudes. Recovering those phases requires additional procedures, such as tomography or interferometric measurements, which may remove the quantum advantage when the full explicit vector is required. For this reason, the HHL is intended to obtain the expectation value of some operator with respect to the solution, or to introduce the solution state into another quantum subroutine.

The main problems of the algorithm are:
\begin{enumerate}
    \item Large amount and waste of resources due to the difference between the size $N$ of the problem and the $n$ qubits to encode it.
    \item Circuit depth and errors introduced by the state preparation and the QPE, in addition to the need for probabilistic post-selection.
    \item $\vec{x}$ is not explicitly obtained from computational-basis sampling alone, and additional measurements are needed to recover signs or relative phases.
    \item The preparation of state $\vec{b}$ may not be trivial, just as performing the inversion operator or performing the $U$ operator.
\end{enumerate}

To try to overcome the first two problems of the HHL, a qudit version of the algorithm is formalized. The main assumption for this algorithm is the existence of quantum computers that implement the basic qudit gates as described in the paper \cite{Qudits}. This assumption is basic, since our work is only theoretical, without going into all the technical difficulties of the hardware. This section is simply a mathematical motivation for the last tensor network algorithm.

The first step is to encode the state $\vec{b}$ in a single qudit. In case the qudit does not have enough states available, the vector is encoded in a number of qudits that allows to encode it in a way analogous to the case of qubits. In the following, it is assumed that only one qudit with $N$ basis states is needed in order to clearly explain the algorithm.

It is important to understand that the goal of this encoding is to reduce the physical amount of quantum resources. It is needed to use a Hilbert space exponentially larger than each of the Hilbert spaces of dimension 2 that would be used in the original case. However, it is assumed that there are physical systems that have this number of states available, since this is the basic premise of the work. An example would be to have a particle with a quantum number with possible values $-3/2,-1/2,1/2,3/2$ instead of another particle with possible values $-1/2,1/2$.

Moreover, this change in the space used for the encoding will not imply a loss of the computational advantage of the original algorithm in the case of being able to realize the gates expressed in the paper \cite{Qudits}, since the unitary gate $U$ is applied over the entire set of state qubits, as if it were a matrix over an ordinary vector.

Now, a way to simulate the $U_{\mu,\tau}=e^{2\pi i \tau A'/\mu}$ operator is needed, which depends on the particular case to be solved. With this, the following circuit in Fig. \ref{fig:HHL Qudits} is performed. With a single qudit of dimension $\mu=2^{n_c}$ the QPE can be performed as in \cite{Qudits} and encode the $\mu$ possible values of the rescaled eigenvalue index in its basis states.
However, more qudits can be used. If 2 qudits are used to encode the $\mu$ values, each one will need to have dimension $\sqrt{\mu}$. This change in resources from clock qubits to a clock qudit will not imply a loss of quantum advantage either, since the algorithm will have the same number of possible eigenvalues.

The inverter is exactly the same as in the case of qubits, but instead of having a control-noncontrol series, there is a control $i$ that applies the rotation gate to the ancilla if there is a value $i$ in the qudit.

The post-selection is performed and if it outputs $|1\rangle$, the inverse QPE is performed to clear the qudit of the eigenvalues.

With this, the number of SWAP gates needed is reduced and the QPE is performed with a low number of gates. In addition, less resources are wasted, as the dimensionality of the quantum system can be better adjusted with respect to the equation to be solved, without the need to have extra elements at 0 to complete the $2^n$ components, which requires an extra qubit. Moreover, in the best-case scenario only two qudits and one qubit are needed, and in the worst-case scenario the same resources as in the original HHL are needed.

The classical simulation of both algorithms can be performed with statevector simulators. However, this simulation can be resource intensive with respect to the size of the problem. In the case of qubits, $n+n_c+1$ qubits are required, so the storage of its statevector has a spatial complexity of $\mathcal{O}\left(2^{(n+n_c)}\right)$, which in terms of $N$ and $\kappa$ is $\mathcal{O}(N\kappa/\epsilon)$. The state preparation usually requires $\mathcal{O}\left(2^{2n}\right) = \mathcal{O}\left(N^2\right)$ operations, due to the application of $\mathcal{O}\left(2^{n}\right)$ dense operations to a dense statevector~\cite{State_Preparation,state_prep}, and then extended for the $n+n_c+1$ qubits with cost $\mathcal{O}\left(2^{(n+n_c)}\right)=\mathcal{O}\left(N\kappa/\epsilon\right)$. However, it can also be initialized directly without other operations and simply extended, reducing the cost of $\mathcal{O}\left(N^2\right)$ to $\mathcal{O}\left(N\right)$. However, the extension will not be applied in this step, to make the QPE in a more efficient way.

The controlled part of QPE requires the application of $n_c$ controlled gates $U^j$. In conventional statevector simulators, which work with dense statevector operations, the $\hat{m}$-th operation would require $\mathcal{O}\left(2^{(2n+\hat{m})}\right)$ operations, because the statevector is extended only to the clock qubits that are involved in every step, so the total $n_c$ steps require
$$\mathcal{O}\left(\sum_{\hat{m}=1}^{n_c}2^{(2n+\hat{m})}\right)=\mathcal{O}\left(2^{(2n+n_c)}\right)=\mathcal{O}\left(N^2\kappa/\epsilon\right)$$
operations~\cite{HHL}. If the $U$ gate is decomposed in $\mathcal{O}\left(\log(N)s\tau\right)$ depth~\cite{hamiltonian_simulation}, requiring $$\mathcal{O}\left(2^{n+n_c}\log(N)s\tau\right)=\mathcal{O}\left(N\kappa/\epsilon\log(N)s\tau\right).$$

Since $\tau=O(\kappa/\epsilon)$~\cite{HHL}, the computational complexity is $\mathcal{O}\left(N\log(N)\kappa^2/\epsilon^2s^2\right)$. However, the inverse QPE does not allow for iterative extension, so it always requires the computation with all the statevector. This means that its computational complexity is $\mathcal{O}\left(2^{(2n+n_c)}n_c\right)=\mathcal{O}\left(N^2\kappa/\epsilon\log(\kappa/\epsilon)\right)$ for the dense $U$ and $\mathcal{O}\left(2^{n+n_c}n_c\log(N)s\tau\right)=\mathcal{O}\left(N\log(N)\kappa^2/\epsilon^2\log(\kappa/\epsilon)s\right)$ for the decomposed.

Inverse QFT requires the application of $\mathcal{O}\left(n_c^2\right)$ operations, so the total computational complexity of this step is $\mathcal{O}\left(2^{(n+n_c)} n_c^2\right)=\mathcal{O}\left(N\kappa/\epsilon \log^2(\kappa/\epsilon)\right)$, negligible with respect to the previous ones.

The inversion circuit can be applied in two main ways. The first is the naive multicontrolled rotations circuit without decomposition, which applies $\mathcal{O}\left(2^{n_c}\right)$ rotations to every possible value of the eigenvalues, controlled by $n_c$ qubits. In conventional statevector simulators, this requires $\mathcal{O}\left(2^{(n+2n_c)}\right)$ operations for each eigenvalue, resulting in $\mathcal{O}\left(2^{(n+3n_c)}\right)=\mathcal{O}\left(N\kappa^3/\epsilon^3\right)$. However, in optimized statevector simulators, the fact that these are controlled operations can be exploited to reduce the complexity per eigenvalue to $\mathcal{O}\left(2^{(n+n_c)}\right)$, resulting in $\mathcal{O}\left(2^{(n+2n_c)}\right)=\mathcal{O}\left(N\kappa^2/\epsilon^2\right)$. The second way consists in the decomposition of each one of these multi-controlled rotations into $\mathcal{O}\left(n_c\right)$ one- and two-qubit gates, resulting in $\mathcal{O}\left(2^{(n+2n_c)}n_c\right)=\mathcal{O}\left(N\kappa^2/\epsilon^2\log(\kappa/\epsilon)\right)$ operations.

All of this implies that the complete computational complexity for simulating the statevector output of the HHL with common statevector simulators has a computational complexity of $$\mathcal{O}\left(N^2\kappa/\epsilon\log(\kappa/\epsilon) + N\kappa^2/\epsilon^2\log(\kappa/\epsilon)\right),$$
neglecting the decomposition and transpilation complexities. With optimized statevector simulators it can be improved to
$$\mathcal{O}\left(N^2\kappa/\epsilon\log(\kappa/\epsilon) + N\kappa^2/\epsilon^2\right).$$

However, all of this process implies the usage of complex decomposition techniques, which incorrectly implemented may increase the computational complexity of the simulation. In regular implementations, without these complex techniques, the computational complexity will remain
$$\mathcal{O}\left(N^2\kappa/\epsilon\log(\kappa/\epsilon) + N\kappa^3/\epsilon^3\right).$$

Still, more can be done to solve the other problems, so the quantum-inspired technique of tensor networks will tackle them, avoiding gate errors from quantum devices and extracting $\vec{x}$, and allowing for a more efficient simulation of the algorithm.

\section{Tensor Networks Algorithm}\label{sec: tensor network}

To return the vector $\vec{x}$ directly, the qudit circuit is transformed into a tensor network. This algorithm is called the \textit{tensor network HHL} (TN HHL). Since in tensor networks normalization is not necessary, the state $|b\rangle$ will not be normalized. As it is not a unit vector, the result state is not normalized either, so it is not necessary to rescale it. Moreover, the state can be prepared exactly in a single operation, defining the node $\vec{b}$ with dimension $N$. In this section, $A$ denotes the Hermitian matrix supplied to phase estimation; in the non-Hermitian case this means the embedded matrix $A'$ of Eq.~\eqref{eq: hermitian}. Accordingly, the phase-estimation unitary is
\begin{equation}
    U=U_{\mu,\tau}=e^{2\pi i \tau A/\mu}.
\end{equation}

The QPE is performed by contracting the uniform superposition clock state with the \textit{Quantum Fourier Transform} (QFT)~\cite{QFT} in the QPE, so it is replaced by a matrix $H[\mu]$ with dimension $\mu\times \mu$ for the $\mu$ eigenvalues with elements
\begin{equation}
    H[\mu]_{ab} = \omega^{ab} = e^{2\pi i\frac{ab}{\mu}},
\end{equation}
where $\omega=e^{2\pi i/\mu}$. This Fourier matrix is intentionally taken \emph{without} the usual $1/\sqrt{\mu}$ normalization, so the two character sums in the tensor-network derivation produce the global factor $\mu^2$.

The inverter is a non-unitary operator with dimension $\mu\times \mu$ for $\mu$ eigenvalues. We write it in terms of the signed inverse map
\begin{equation}
    g_{\mu}(d) = \left\{
    \begin{array}{cl}
        0 & \mbox{if } d=0, \\
        \nicefrac{1}{d} & \mbox{if } 1\leq d\leq \lfloor \mu/2 \rfloor, \\
        \nicefrac{1}{d-\mu} & \mbox{if } \lfloor \mu/2 \rfloor < d \leq \mu-1,
    \end{array}
\right.
\qquad
\mbox{inv}[\mu]_{i,j} = \delta_{ij} g_{\mu}(i).
\end{equation}
The index $d=0$ is not invertible and is therefore assigned the value $0$. Positive and negative eigenvalue estimates are represented cyclically: for $d>\mu/2$ the signed representative is $d-\mu$. When $\mu$ is even we adopt the convention that the Nyquist index $d=\mu/2$ is treated as positive, so $g_{\mu}(\mu/2)=2/\mu$. This convention matches the tensor formulas used below. If $|\tau \lambda|\geq \mu/2$, aliasing occurs because different signed representatives collapse to the same clock value; moreover, eigenvalues close to zero are unstable because the inverse map is singular there.

The phase kickback operators can also be obtained exactly from $U$. This tensor $P$ with dimension $N\times \mu\times N$ is
\begin{equation}
    P[\mu]_{i,j,k} = \left(U^j\right)_{i,k}; \left(P[\mu]^{-1}\right)_{i,j,k} = \left(\left(U^{-1}\right)^j\right)_{k,i}.
\end{equation}
These tensors are contracted through their $j$ index with the $H[\mu]$ and $H[\mu]^{-1}$ tensors to perform QPE.

\begin{figure}[h]
  \centering
\includegraphics[width=\linewidth]{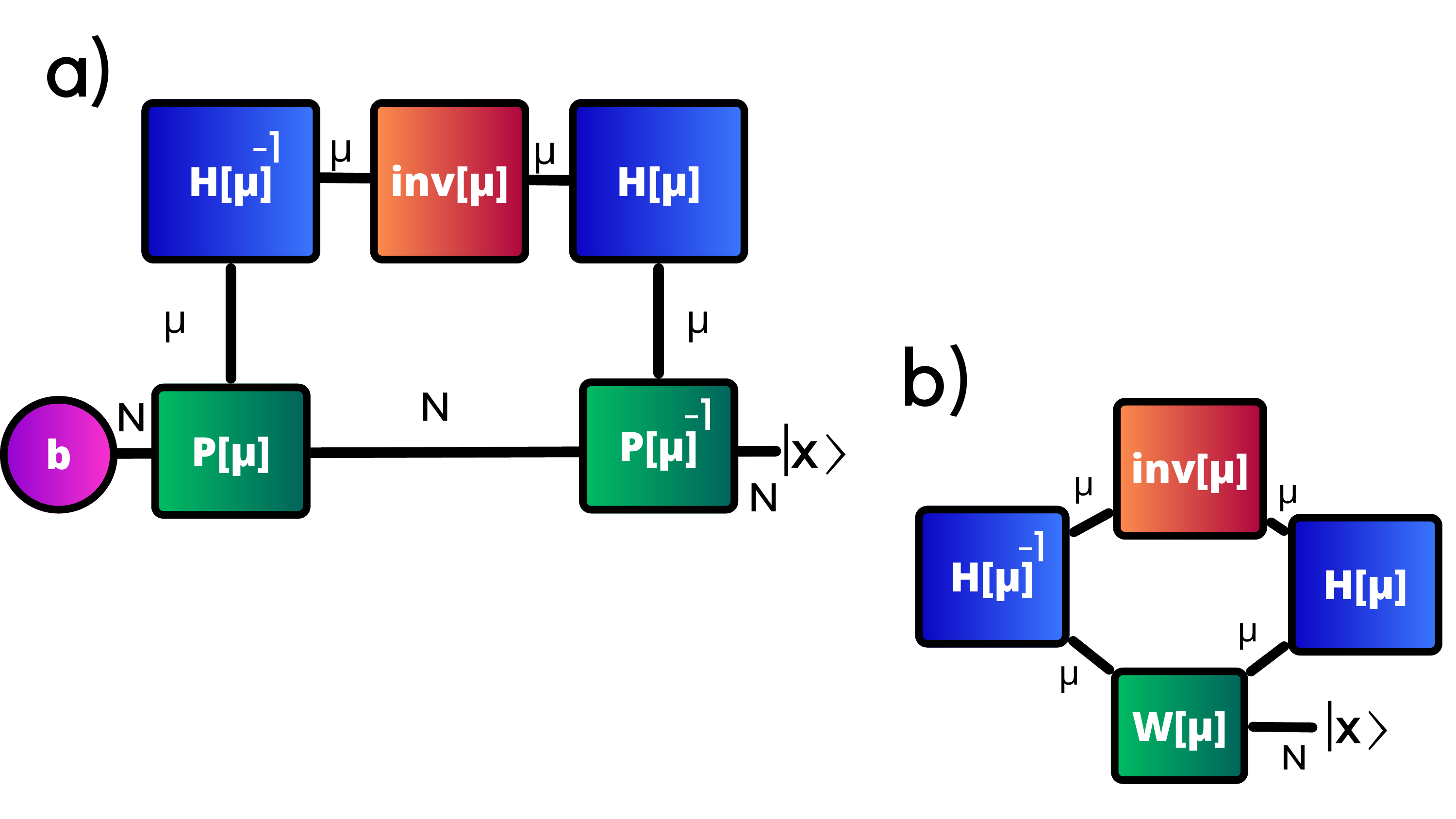}
  \caption{Tensor network equivalent to HHL. a) Original way. b) Efficient way.}
  \label{fig:Tensor Network HHL}
\end{figure}

With these tensors, the contraction of the tensor network in Fig. \ref{fig:Tensor Network HHL} a defines the operator
\begin{equation}
    \left(T_{\mu,\tau}\vec{b}\right)_i :=
    \sum_{a,b,c,d,e,f} b_a P_{abc} H^{-1}_{bd} \mbox{inv}_{de} H_{ef} P^{-1}_{cfi},
    \label{eq:tn_tensor_operator}
\end{equation}
omitting the label $[\mu]$ in the tensor names. Equivalently,
\begin{equation}
    T_{\mu,\tau} =
    \sum_{b,d,f=0}^{\mu-1}
    \omega^{-bd} g_{\mu}(d) \omega^{df} U^b U^{-f}.
    \label{eq:tn_operator_sum}
\end{equation}

\begin{proposition}\label{prop:tn-exact}
Let $A$ be Hermitian and invertible, with spectral decomposition $A=\sum_r \lambda_r \ket{u_r}\!\bra{u_r}$, and let $\vec{b}=\sum_r \beta_r \ket{u_r}$. Assume that for every $r$ with $\beta_r\neq 0$ there exists a non-zero integer $d_r$ such that
\begin{equation}
    d_r=\tau \lambda_r,
    \qquad
    |d_r|<\mu/2,
\end{equation}
with $d_r$ represented modulo $\mu$. Then
\begin{equation}
    T_{\mu,\tau}\vec{b} = \frac{\mu^2}{\tau} A^{-1}\vec{b}.
    \label{eq:tn_exact_result}
\end{equation}
\end{proposition}

\begin{proof}
Acting on an eigenvector $\ket{u_r}$, one has $U^b\ket{u_r}=\omega^{bd_r}\ket{u_r}$. Therefore
\begin{align}
    T_{\mu,\tau}\ket{u_r}
    &= \sum_{d=0}^{\mu-1} g_{\mu}(d)
    \left(\sum_{b=0}^{\mu-1}\omega^{b(d_r-d)}\right)
    \left(\sum_{f=0}^{\mu-1}\omega^{f(d-d_r)}\right)\ket{u_r} \nonumber \\
    &= \mu^2 g_{\mu}(d_r)\ket{u_r},
\end{align}
where the character orthogonality relation
\begin{equation}
    \sum_{b=0}^{\mu-1}\omega^{b(d_r-d)} = \mu \,\delta_{d,\, d_r \, \mathrm{mod}\, \mu}
\end{equation}
has been used, and similarly for the sum over $f$. Since $d_r=\tau \lambda_r$ and $|d_r|<\mu/2$, the signed representative selected by $g_{\mu}$ is exactly $d_r$, so
\begin{equation}
    T_{\mu,\tau}\ket{u_r}
    = \frac{\mu^2}{\tau \lambda_r}\ket{u_r}.
\end{equation}
Extending by linearity to $\vec{b}=\sum_r \beta_r \ket{u_r}$ proves Eq.~\eqref{eq:tn_exact_result}.
\end{proof}

\begin{proposition}\label{prop:tn-filter}
For general eigenvalues, define
\begin{equation}
    D_{\mu}(z)=\sum_{b=0}^{\mu-1} e^{2\pi i b z/\mu}
\end{equation}
and
\begin{equation}
    f_{\mu,\tau}(\lambda)
    =
    \frac{\tau}{\mu^2}
    \sum_{d=0}^{\mu-1}
    g_{\mu}(d)\left|D_{\mu}(\tau\lambda-d)\right|^2.
\end{equation}
Then the rescaled tensor-network output satisfies
\begin{equation}
    \frac{\tau}{\mu^2} T_{\mu,\tau}\vec{b} = f_{\mu,\tau}(A)\vec{b},
    \label{eq:tn_filter_result}
\end{equation}
and therefore
\begin{equation}
    \left\|f_{\mu,\tau}(A)\vec{b}-A^{-1}\vec{b}\right\|
    \leq
    \max_{\lambda\in\operatorname{spec}(A)}
    \left|f_{\mu,\tau}(\lambda)-\frac{1}{\lambda}\right|
    \|\vec{b}\|.
    \label{eq:tn_filter_bound}
\end{equation}
\end{proposition}

\begin{proof}
For an eigenvector $\ket{u_r}$ with eigenvalue $\lambda_r$,
\begin{equation}
    T_{\mu,\tau}\ket{u_r}
    =
    \sum_{d=0}^{\mu-1}
    g_{\mu}(d)\left|D_{\mu}(\tau\lambda_r-d)\right|^2\ket{u_r}
    =
    \frac{\mu^2}{\tau}f_{\mu,\tau}(\lambda_r)\ket{u_r}.
\end{equation}
The spectral theorem then gives Eq.~\eqref{eq:tn_filter_result}. Equation~\eqref{eq:tn_filter_bound} follows from the operator norm bound for a Hermitian matrix diagonalized in its eigenbasis.
\end{proof}

\begin{remark}\label{rem:tn_assumptions}
Therefore, equalities of the form $\left(T_{\mu,\tau}\vec{b}\right)_i=\mu^2 x_i/\tau$ should be read as exact only under the phase-grid and no-aliasing hypotheses of Proposition~\ref{prop:tn-exact}. In general the tensor network applies the spectral filter $f_{\mu,\tau}(A)$ of Proposition~\ref{prop:tn-filter}, not the exact inverse. Consequently, statements such as $\mu=O(\kappa/\epsilon)$ require assumptions controlling the spectral approximation error in Eq.~\eqref{eq:tn_filter_bound}, a window $0<\lambda_{min}\leq |\lambda_r|$, the no-aliasing condition $|\tau\lambda_r|<\mu/2$, and separation from the singular bin $d=0$.
\end{remark}

It is assumed that contracting a tensor of $N$ indexes of dimension $n_i$ with another of $M$ indexes of dimension $m_j$ through its first index has the usual computational cost of $\mathcal{O}\left(n_0\prod_{i,j=1,1}^{N-1,M-1} n_i m_j\right)$. 

The computational complexity of contracting the tensors $H[\mu]^{-1}$ and $\mbox{inv}[\mu]$ is $\mathcal{O}\left(\mu^2\right)$, due to the sparsity of the tensor $\mbox{inv}[\mu]$. Contracting the resulting tensor with $H[\mu]$ has a complexity of $\mathcal{O}\left(\mu^3\right)$. The contraction of $\vec{b}$ with $P[\mu]$ has a complexity of $\mathcal{O}\left(N^2\mu \right)$. The contraction of the tensors resulting from the contraction of $H[\mu]^{-1}$, $H[\mu]$ and $\mbox{inv}[\mu]$ and the contraction of $\vec{b}$ and $P[\mu]$ has a computational complexity of $\mathcal{O}\left(N\mu^2\right)$. And the final contraction with $P[\mu]^{-1}$ has a complexity of $\mathcal{O}\left(N^2\mu \right)$.

Therefore, the contraction of the tensor network has a computational complexity of\\$\mathcal{O}\left(N^2\mu+ N\mu^2 +\mu^3\right)$.

The construction of the tensors has a complexity:
\begin{itemize}
    \item $H[\mu]^{-1}$ and $H[\mu]$: $\mathcal{O}\left(\mu^2\right)$.
    \item $\mbox{inv}[\mu]$: $\mathcal{O}\left(\mu\right)$.
    \item $\vec{b}$: $\mathcal{O}\left(N\right)$.
    \item $P[\mu]$ and $P[\mu]^{-1}$: $\mathcal{O}\left(\mu N^3\right)$, since $N\times N$ matrices have to be multiplied up to $\mu$ times.
\end{itemize}
Therefore, the computational complexity of the algorithm is $\mathcal{O}\left(N^3\mu+N\mu^2 +\mu^3\right)$.

The cost of constructing or applying $U_{\mu,\tau}$ depends on the structure of $A$. If $U_{\mu,\tau}$ is formed explicitly from a dense matrix, standard matrix-exponential routines can cost $\mathcal{O}(N^3)$ and may dominate when $\mu$ is small compared with $N$. The contraction complexities quoted below therefore assume either that $U_{\mu,\tau}$ is supplied as an oracle or precomputed matrix, or that $A$ has enough structure to apply $U_{\mu,\tau}$ efficiently; otherwise this preprocessing cost must be added separately. The same caveat applies to $U_{\mu,\tau}^{-1}$.

This increase in complexity due to the construction of the tensors $P[\mu]$ and $P[\mu]^{-1}$ can be avoided by defining a tensor $W[\mu]$ that directly computes the contraction of both tensors with the vector $\vec{b}$. Its elements are
\begin{equation}
    W[\mu]_{ijk} = \left(\vec{b}U^{i-j}\right)_k.
\end{equation}
This tensor has dimension $\mu\times \mu\times N$, but it depends only on the difference $i-j$. Therefore, only $2\mu-1$ distinct vectors
\begin{equation}
    \vec{v}_{\ell}=\vec{b}U^{\ell},
    \qquad
    \ell=-(\mu-1),\ldots,\mu-1,
\end{equation}
are needed. If these vectors are generated recursively, the algebraic cost is $\mathcal{O}\left(\mu N^2\right)$ for dense $U$. Explicitly materializing all entries of $W[\mu]$ from them requires additional time and memory $\mathcal{O}\left(N\mu^2\right)$, with an unavoidable writing cost $\Omega\left(N\mu^2\right)$. Therefore, the explicit construction cost is $\mathcal{O}\left(\mu N^2 + N\mu^2\right)$, whereas the smaller $\mathcal{O}\left(\mu N^2\right)$ figure corresponds to an implicit representation that stores only the vectors $\vec{v}_{\ell}$.

So, the tensor network in Fig. \ref{fig:Tensor Network HHL} b has to be contracted, representing the equation
\begin{equation}
    \left(T_{\mu,\tau}\vec{b}\right)_i
    =
    \sum_{a,b,c,d} W_{abi} H^{-1}_{ac} \mbox{inv}_{cd} H_{db}.
    \label{eq:tn_w_operator}
\end{equation}
Equation~\eqref{eq:tn_w_operator} represents the same operator as Eq.~\eqref{eq:tn_tensor_operator}, so Proposition~\ref{prop:tn-exact} and Proposition~\ref{prop:tn-filter} apply to this version as well. The complexity of contracting tensors $H[\mu]^{-1}$, $H[\mu]$, and $\mbox{inv}[\mu]$ is the same as before. Contracting this resulting tensor with $W[\mu]$ has a complexity of $\mathcal{O}\left(N\mu^2\right)$. Therefore, the explicit total computational complexity is $\mathcal{O}\left(\mu N^2+N\mu^2 +\mu^3\right)$, plus any preprocessing cost needed to form $U_{\mu,\tau}$. If the contraction of $H[\mu]$, $H[\mu]^{-1}$ and $inv[\mu]$ is pre-calculated to be used every time, the $\mathcal{O}(\mu^3)$ term could be avoided.

Moreover, it can be slightly improved in the case of $\mu=2^n_c$ contracting the $W[\mu]$ tensor with the QFT circuits instead of dense tensors. This contraction consists of $n_c^2$ contractions of a tensor with $\mathcal{O}\left(2^{(n+2n_c)}\right)$ elements with a tensor of one or two qubits, with a resulting complexity of $\mathcal{O}\left(N\mu^2  \log^2(\mu)\right)$. Then, it is contracted with the $inv[\mu]$ tensor with complexity $\mathcal{O}\left(2^{(n+2n_c)}\right)$ because it is diagonal, and finally with the other QFT circuit. This makes the algorithm require only $\mathcal{O}\left(N^2\mu+N\mu^2  \log^2(\mu)\right)$ operations, avoiding the cubic scaling. This is the TN HHL W with QFT. Another possibility is to first multiply $inv[\mu]$ with the two QFT circuits, each with complexity $\mathcal{O}\left(2^{2n_c}n^2_c\right)=\mathcal{O}\left(\mu^2  \log^2(\mu)\right)$, and a final contraction with $W[\mu]$ with complexity $\mathcal{O}\left(2^{(n+2n_c)}\right)=\mathcal{O}\left(N\mu^2\right)$. Then, the total complexity is $\mathcal{O}\left(N^2\mu+N\mu^2+\mu^2\log^2(\mu)\right)$, again avoiding the cubic term. This is the TN HHL inv with QFT.

The explicit spatial complexity is $\mathcal{O}(N\mu^2+N^2)$, being the first term associated with the tensor $W[\mu]$ and the second term associated with the matrix $U$. If $W[\mu]$ is stored implicitly through the vectors $\vec{v}_{\ell}$, this reduces to $\mathcal{O}(N\mu+N^2)$.

The inverse of $A$ can also be approximated by erasing the $b$ node from Fig. \ref{fig:Tensor Network HHL} a and performing the contraction, increasing by a factor of $N$ the explicit complexities above. Under Proposition~\ref{prop:tn-exact} this returns $(\mu^2/\tau)A^{-1}$ on the no-aliasing phase grid; in the general case it returns the filtered operator associated with Proposition~\ref{prop:tn-filter}.

\section{Comparison of advantages and disadvantages}\label{sec: comparison}
The computational complexities of these algorithms in tensor networks are compared in Table \ref{tab: times} against the relevant classical solvers CG and LU, the quantum HHL and the state-vector simulations.

\begin{table*}[ht]
\centering
\begin{tabular}{|l|l|l|l}
\cline{1-3}
\textbf{Algorithm} & \textbf{Solution $\vec{x}=A^{-1}\vec{b}$} & \textbf{Expectation value $\vec{x^T}M\vec{x}$}    &  \\ \cline{1-3}
CG      & $\mathcal{O}(Ns\sqrt{\kappa}\log(1/\epsilon))$         & $\mathcal{O}(Ns\sqrt{\kappa}\log(1/\epsilon) + Ns')$               &  \\ \cline{1-3}
LU        & $\mathcal{O}(N^3)$         & $\mathcal{O}(N^3)$               &  \\ \cline{1-3}
HHL        & -                      & $\mathcal{O}(\log(N)s^2\kappa^2/\epsilon^3)$ &  \\ \cline{1-3}
Naive Statevector       & $\mathcal{O}\left(N^2\left(\frac{\kappa}{\epsilon}\right)\log\left(\frac{\kappa}{\epsilon}\right) + N\left(\frac{\kappa}{\epsilon}\right)^3\right)$ & $\mathcal{O}\left(N^2\left(\frac{\kappa}{\epsilon}\right)\log\left(\frac{\kappa}{\epsilon}\right) + N\left(\frac{\kappa}{\epsilon}\right)^3\right)$ &  \\ \cline{1-3}
Common Statevector       & $\mathcal{O}\left(\left(N^2\left(\frac{\kappa}{\epsilon}\right) + N\left(\frac{\kappa}{\epsilon}\right)^2\right)\log\left(\frac{\kappa}{\epsilon}\right)\right)$ & $\mathcal{O}\left(\left(N^2\left(\frac{\kappa}{\epsilon}\right) + N\left(\frac{\kappa}{\epsilon}\right)^2\right)\log\left(\frac{\kappa}{\epsilon}\right)\right)$ &  \\ \cline{1-3}
Optimized Statevector       & $\mathcal{O}\left(N^2\left(\frac{\kappa}{\epsilon}\right)\log\left(\frac{\kappa}{\epsilon}\right) + N\left(\frac{\kappa}{\epsilon}\right)^2\right)$ & $\mathcal{O}\left(N^2\left(\frac{\kappa}{\epsilon}\right)\log\left(\frac{\kappa}{\epsilon}\right) + N\left(\frac{\kappa}{\epsilon}\right)^2\right)$ &  \\ \cline{1-3}
\textbf{TN HHL}   & $\mathcal{O}\left(N^2\left(\frac{\kappa}{\epsilon}\right)+N\left(\frac{\kappa}{\epsilon}\right)^2+\left(\frac{\kappa}{\epsilon}\right)^3\right)$          & $\mathcal{O}\left(N^2\left(\frac{\kappa}{\epsilon}\right)+ N\left(\frac{\kappa}{\epsilon}\right)^2+ \left(\frac{\kappa}{\epsilon}\right)^3\right)$                  &  \\ \cline{1-3}
\textbf{TN HHL W with QFT}   & $\mathcal{O}\left(N^2\left(\frac{\kappa}{\epsilon}\right)+N\left(\frac{\kappa}{\epsilon}\right)^2  \log^2\left(\frac{\kappa}{\epsilon}\right)\right)$          & $\mathcal{O}\left(N^2\left(\frac{\kappa}{\epsilon}\right)+N\left(\frac{\kappa}{\epsilon}\right)^2  \log^2\left(\frac{\kappa}{\epsilon}\right)\right)$                  &  \\ \cline{1-3}
\textbf{TN HHL inv with QFT}   & $\mathcal{O}\left(N^2\left(\frac{\kappa}{\epsilon}\right)+N\left(\frac{\kappa}{\epsilon}\right)^2+\left(\frac{\kappa}{\epsilon}\right)^2\log^2\left(\frac{\kappa}{\epsilon}\right)\right)$          & $\mathcal{O}\left(N^2\left(\frac{\kappa}{\epsilon}\right)+N\left(\frac{\kappa}{\epsilon}\right)^2+\left(\frac{\kappa}{\epsilon}\right)^2\log^2\left(\frac{\kappa}{\epsilon}\right)\right)$                   &  \\ \cline{1-3}
\end{tabular}
\caption{Computational times to obtain the solution of $A\vec{x}=\vec{b}$ and compute an expectation value $\langle x | M | x \rangle$. $s$ and $s'$ are the maximum number of non-zero elements per row of the $N\times N$ matrices $A$ and $M$. The CG row uses the standard sparse Hermitian positive definite complexity estimate.}
\label{tab: times}
\end{table*}

\subsection{Quantum, statevector and TN algorithms vs classical algorithms}
The quantum HHL algorithm is faster in terms of $N$ than the other algorithms for the computation of the expectation value, but it is not an algorithm to extract the explicit solution. In the sparse positive definite regime where CG applies, the HHL statevector simulation is significantly slower than the CG, but it is faster than the LU in situations where $\kappa/\epsilon$ is small enough. With more precision, if $\left(\frac{\kappa}{\epsilon}\right)\log\left(\frac{\kappa}{\epsilon}\right)<N$. The TN HHL algorithms are also slower than the CG, but faster than the LU if $\left(\frac{\kappa}{\epsilon}\right)<N$. These are reasonable situations because it makes more sense to use the HHL algorithm in situations where $\kappa/\epsilon\approx\mathcal{O}\left(\mathrm{polylog}(N)\right)$~\cite{HHL}. However, the multiplicative constant in complexity keeps these algorithms slower than the LU solver.

\subsection{Quantum HHL vs Statevector HHL vs TN HHL}
The quantum HHL is much faster than its classical simulations for the expectation value computation. The faster statevector method is the Optimized, and the faster TN method is the TN HHL inv with QFT. Both algorithms have similar complexities, but the TN is faster if $\left(\frac{\kappa}{\epsilon}\right)\log\left(\frac{\kappa}{\epsilon}\right)<N^2$, which is a reasonable assumption. If the comparison is between the most naive simulators, the TN HHL is always faster than the naive statevector and is faster than the common statevector if $\left(\frac{\kappa}{\epsilon}\right)/\sqrt{\log\left(\frac{\kappa}{\epsilon}\right)}<N$, reasonable again.

\section{Experiments}\label{sec: simulation}

The effectiveness of the TN HHL algorithm is tested by solving three common simple problems with the expected structure of the problems that are the target of the quantum HHL algorithm. The forced harmonic oscillator, the forced damped oscillator, and the 2D static heat equation with sources are solved and compared against the solution provided by the PyTorch solver, which makes use of the LU decomposition algorithm. Then, the performance of the TN HHL is compared against a Qiskit implementation of the naive statevector HHL for random sparse matrices. Finally, the sensitivity of the quality of the HHL solution is studied with respect to their hyperparameter values, making use of the TN HHL.

It is important to note that the comparison is about the resolution with respect to the system of linear equations, not with respect to the original differential equation itself. This is because this method solves the discretization of the problem, so errors due to the discretization itself should not be taken into account. In each experiment, the hyperparameters are tuned by hand.

All experiments are performed in CPU, with an Intel(R) Core(TM) i7-14700HX 2.10 GHz and 16 GB RAM.
All experiments reported in this section are real-valued. For general complex-valued systems, both the implementation and the error metrics must retain the full complex output rather than only its real part, together with a consistent phase-alignment convention when explicit vectors are compared.

\subsection{Forced harmonic oscillator}\label{ssec: forced}
This is the most simple problem with a complex structure. The differential equation to be solved is
\begin{gather}
    \frac{d^2x}{dt^2} + \frac{k}{m}x = F(t)\\
    x_0 = x(t=0); \qquad x_T = x(t=T) \nonumber
\end{gather}
where $F(t)$ is the external force dependent on time $t$. For experiments, a harmonic force $F(t)=9 \sin(0.4 t)$ is chosen to be strong enough to modify the evolution of the system and with a frequency that does not synchronize with the system frequency.
\begin{figure}
  \centering
  \includegraphics[width=\linewidth]{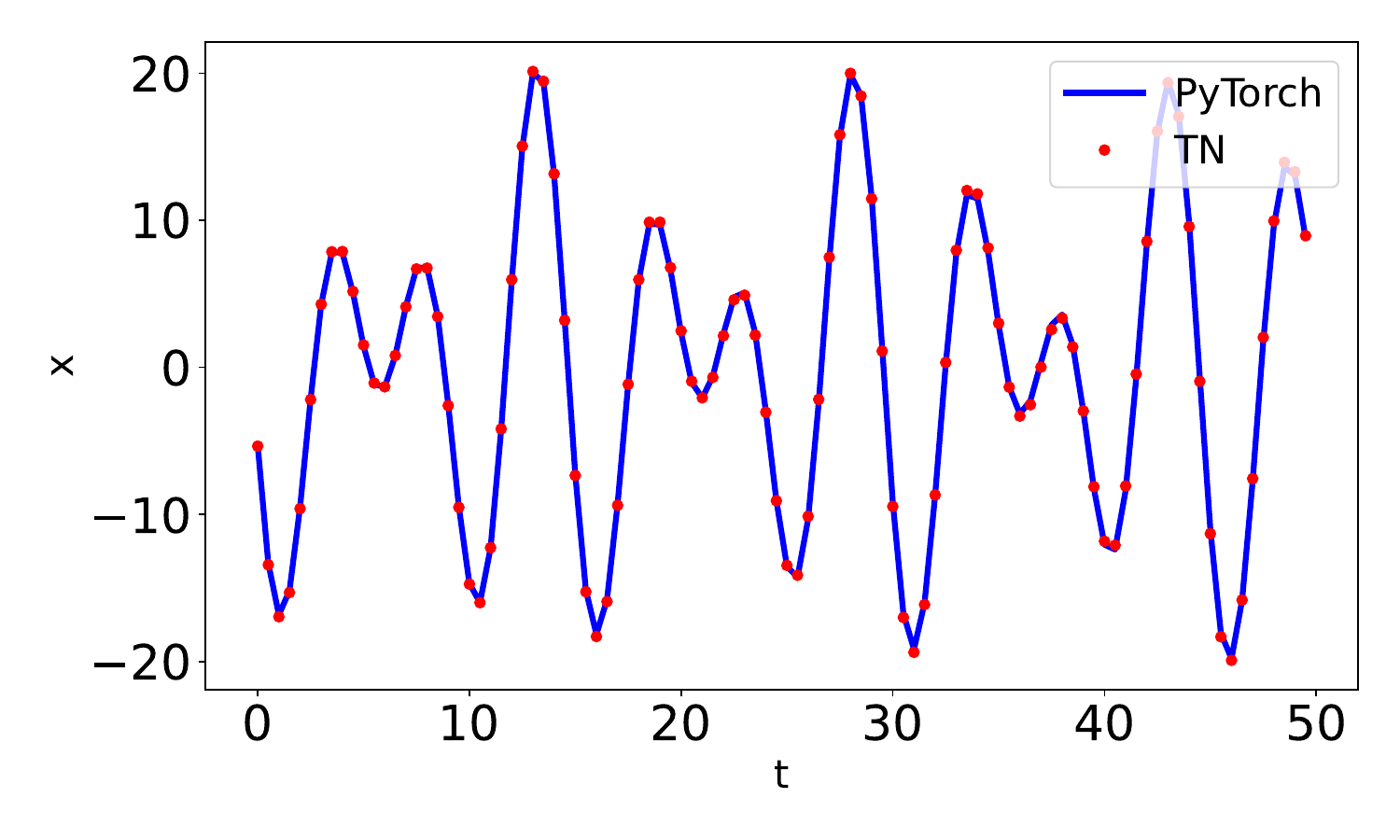}
  \caption{Solving the forced harmonic oscillator system with equation \eqref{eq: lineal OAF}. In blue the inversion performed with PyTorch and in red the inversion performed with the tensor network. The parameters are $k=5, m=7, x_0=5, x_T=3, \Delta t=0.5, T=50$.}
  \label{fig:OAF}
\end{figure}

A discretization with $n$ time steps is applied, i.e. $\Omega=-2+\frac{k}{m}(\Delta t)^2$ and $F_j=(\Delta t)^2 9 \sin(0.4 j \Delta t)$.
\begin{equation}
    \left(
    \begin{matrix}
    \Omega & 1 & 0 & \cdots & 0 & 0 \\
    1 & \Omega & 1 & \cdots & 0 & 0 \\
    0 & 1 & \Omega & \cdots & 0 & 0 \\
    \vdots & \vdots & \vdots & \ddots & \vdots & \vdots \\
    0 & 0 & 0 & \cdots & 1 & \Omega
    \end{matrix}
    \right)
    \left(
    \begin{matrix}
    x_1 \\
    x_2 \\
    x_3 \\
    \vdots \\
    x_n
    \end{matrix}
    \right)=
    \left(
    \begin{matrix}
    F_1-x_0 \\
    F_2 \\
    F_3 \\
    \vdots \\
    F_n - x_T
    \end{matrix}
    \right)
    \label{eq: lineal OAF}
\end{equation}
The result of inverting this system gives us the result in Fig. \ref{fig:OAF}. 
As hyperparameters of the algorithm of the method, $\mu=2000$ and $\tau=6000$ are chosen.

The root mean square error of our tensor network from the exact inversion was $1.8\times 10^{-5}$ and took $407$ ms to run, compared to $244$ $\mu$s of the exact inversion method of PyTorch.

\subsection{Forced damped oscillator}
This is a slightly more complex problem than the previous one. The differential equation to be solved is
\begin{gather}
    \frac{d^2x}{dt^2} + \gamma \frac{dx}{dt} + \frac{k}{m}x = F(t)\\
    x_0 = x(t=0); \qquad x_T = x(t=T)\nonumber
\end{gather}
where $F(t)$ is the external force dependent on time and $\gamma$ is the damp coefficient. As in Sec. \ref{ssec: forced}, for the experiments, a harmonic force $F(t)=9\sin(0.4 t)$ is chosen.

\begin{figure}
  \centering
  \includegraphics[width=\linewidth]{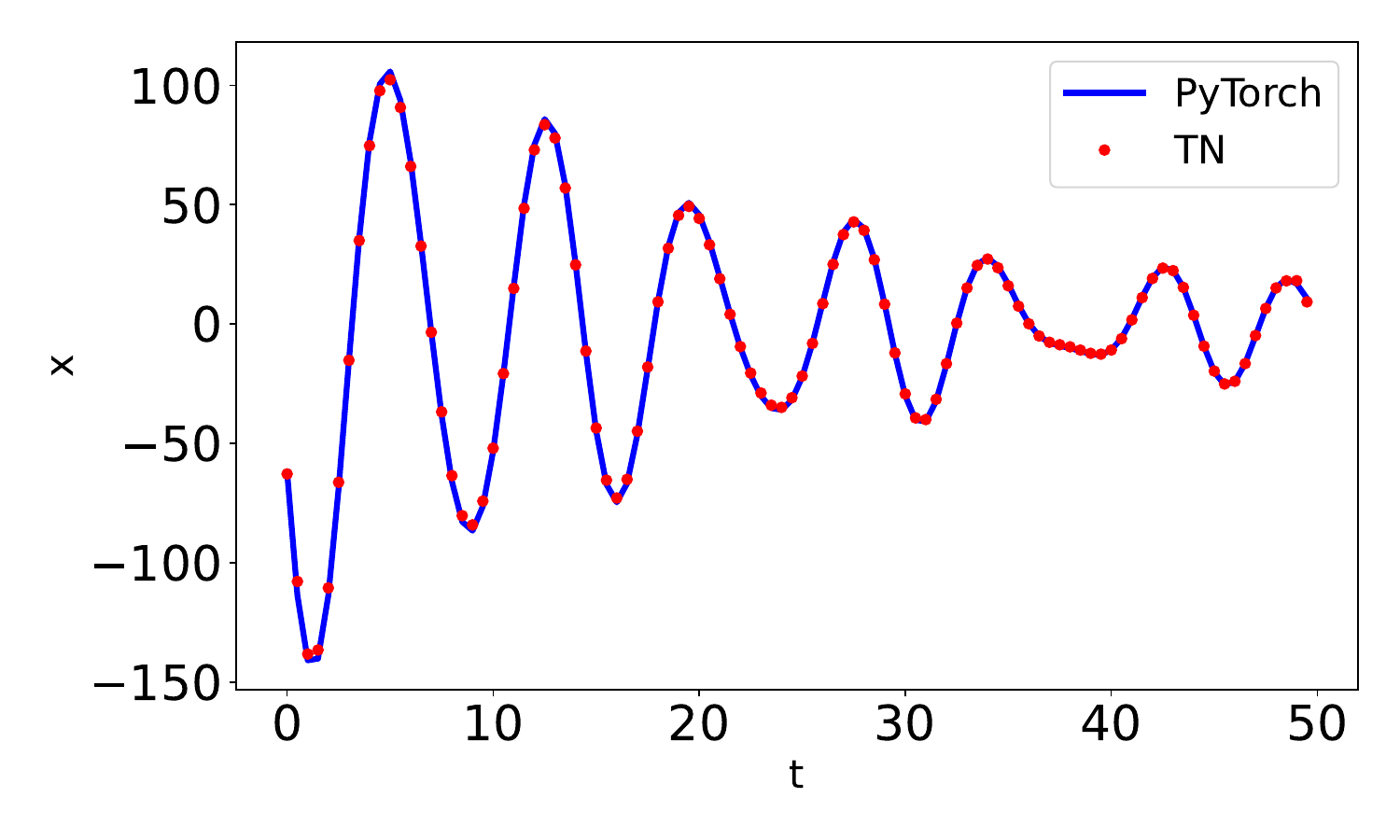}
  \caption{Solving the forced damped oscillator system with equation \eqref{eq: lineal OAAF}. In blue the inversion performed with PyTorch and in red the inversion performed with the tensor network. The parameters are $k=5, m=7, \gamma=0.1, x_0=5, x_T=3, \Delta t=0.5, T=50$.}
  \label{fig:OAAF}
\end{figure}

A discretization with $n$ time steps is applied
\begin{equation}
    \left(
    \begin{matrix}
    \beta_0 & \beta_+ & 0 & \cdots & 0 & 0 \\
    \beta_- & \beta_0 & \beta_+ & \cdots & 0 & 0 \\
    0 & \beta_- & \beta_0 & \cdots & 0 & 0 \\
    \vdots & \vdots & \vdots & \ddots & \vdots & \vdots \\
    0 & 0 & 0 & \cdots & \beta_- & \beta_0
    \end{matrix}
    \right)
    \left(
    \begin{matrix}
    x_1 \\
    x_2 \\
    x_3 \\
    \vdots \\
    x_n
    \end{matrix}
    \right)=
    \left(
    \begin{matrix}
    F_1-\beta_-x_0 \\
    F_2 \\
    F_3 \\
    \vdots \\
    F_n - \beta_+x_T
    \end{matrix}
    \right)
    \label{eq: lineal OAAF}
\end{equation}
where $\beta_-=1-\gamma \frac{\Delta t}{2}$, $\beta_+=1+\gamma \frac{\Delta t}{2}$ and $\beta_0=-2+\frac{k}{m}(\Delta t)^2$.
This matrix is not Hermitian, so \eqref{eq: hermitian} is applied and \eqref{eq: non hermitian} is the equation to solve.

The result of inverting this matrix gives us the result in Fig. \ref{fig:OAAF}. 
As hyperparameters of the algorithm of the method, $\mu=2000$ and $\tau=1.1\times 10^4$ are chosen.

The relative root mean square error of our tensor network from the exact inversion was $6.1\times 10^{-3}$ and took $1.19$ seconds to run, compared to $167$ $\mu$s of the PyTorch method.

\subsection{Static two dimensional heat equation with sources}
This is a more complex problem than the two previous ones because of its bidimensional structure and interactions. The differential equation to be solved is
\begin{gather}
    k\left(\frac{d^2u}{dx^2} + \frac{d^2u}{dy^2} \right) = -S(x,y)\\
    u_{x1} = u(0, y); \quad u_{x2} = u(L_x, y)\nonumber\\
    u_{y1} = u(x, 0); \quad u_{y2} = u(x, L_y)\nonumber
\end{gather}
where $S(x,y)$ is the external source dependent on position. For the experiments, a source $S(x,y)= 10 \sin\left(2\pi \frac{xy}{L_x L_y}\right)$ is chosen to influence the system but not saturate its dynamics.

\begin{figure}
    \centering
    \begin{subfigure}[t]{0.48\textwidth}
        \centering
        \includegraphics[width=\linewidth]{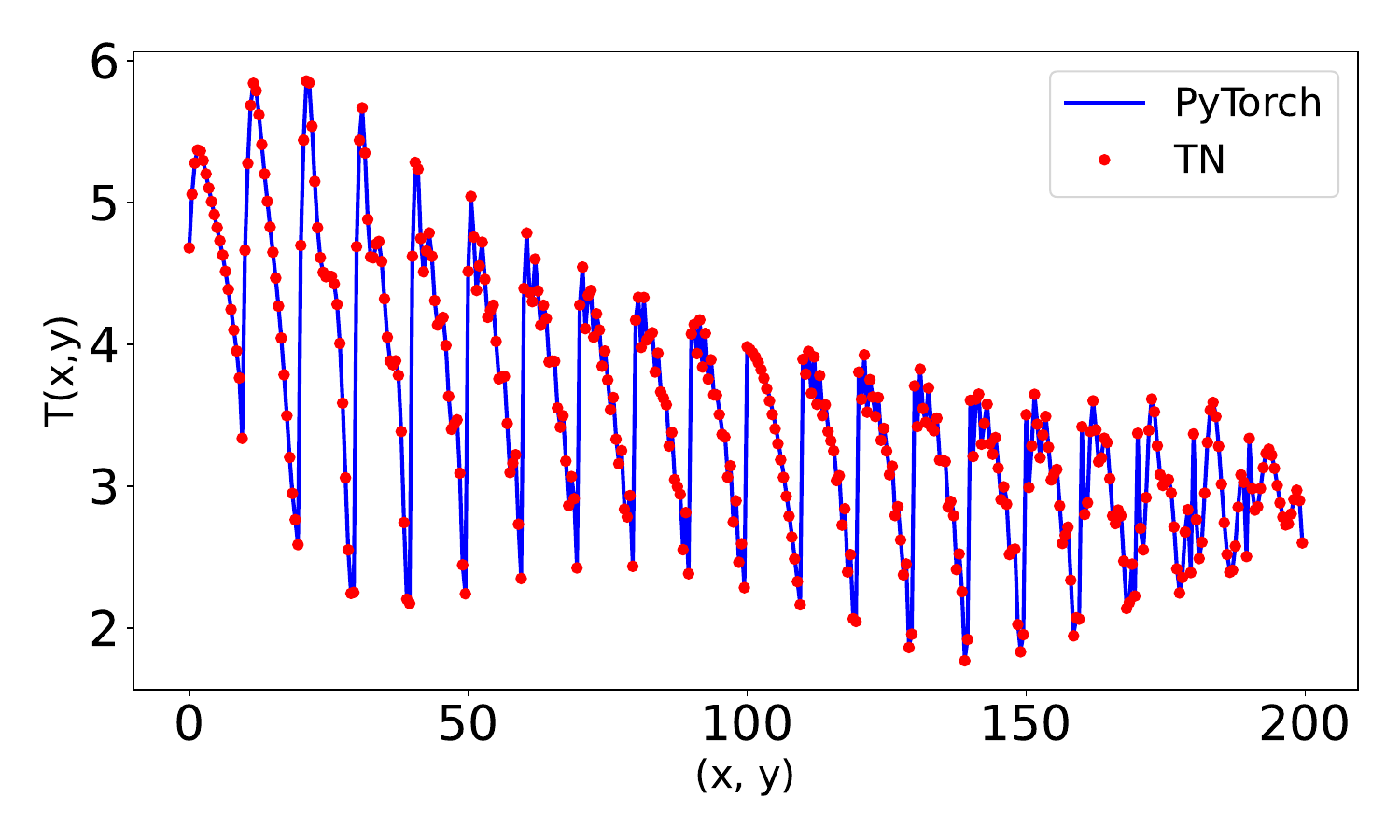}
        \caption{Flattened view of the solution. In blue the inversion performed with PyTorch and in red the inversion performed with the tensor network.}
        \label{fig:C2D}
    \end{subfigure}
    \begin{subfigure}[t]{0.48\textwidth}
        \centering
        \includegraphics[width=\linewidth]{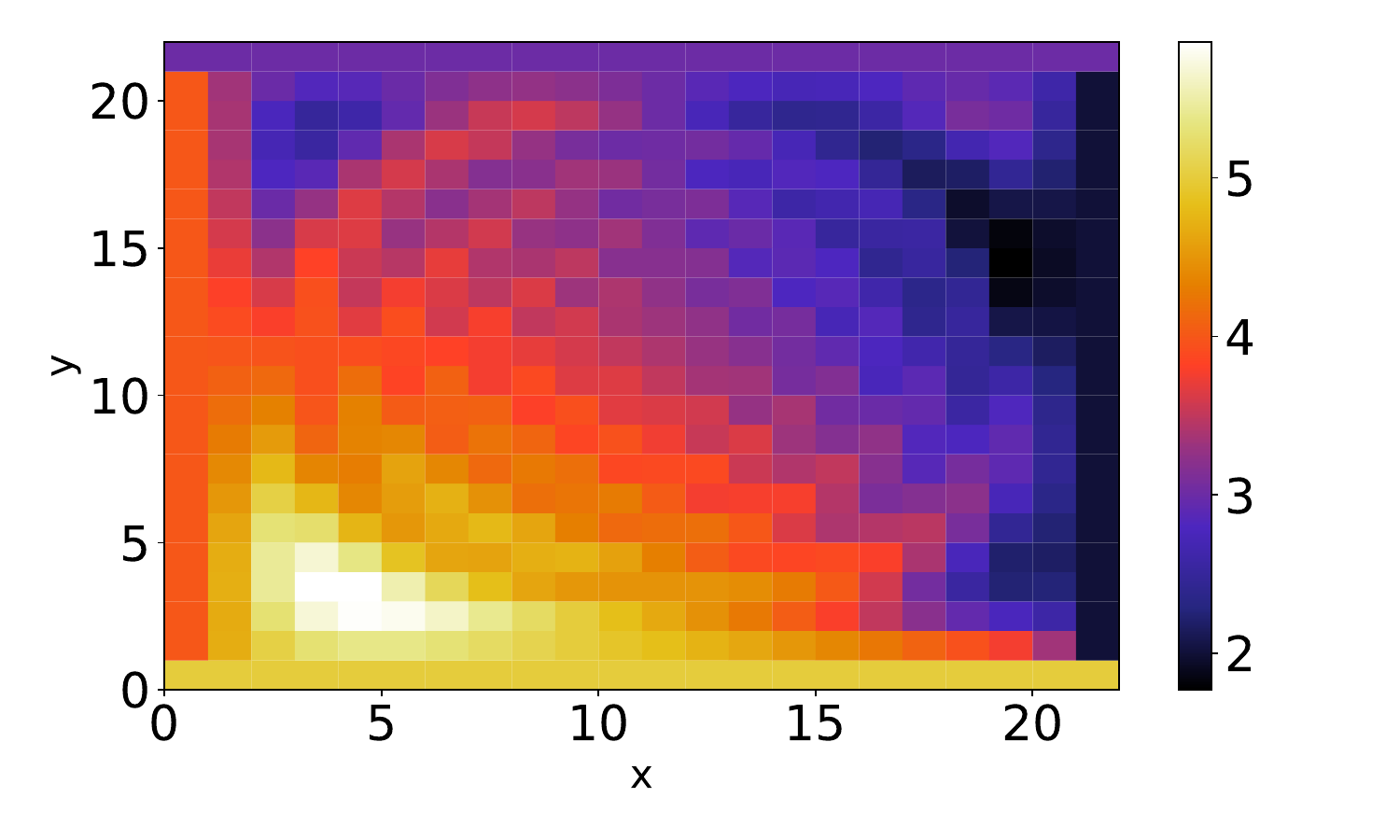}
        \caption{Bidimensional view of the solution.}
        \label{fig:C2D_2D}
    \end{subfigure}
    \caption{Solving the static two dimensional heat equation with sources with equation \eqref{eq: C2D}. The parameters are $k=3, u_{x1}=5, u_{x2}=3, u_{y1}=4, u_{y2}=2, \Delta x = 0.5, L_x= 10, L_y = 10$.}
    \label{fig:C2D_combined}
\end{figure}

The chosen discretization is
\begin{equation}
    u_{j+1,k} + u_{j-1,k} + u_{j,k+1} + u_{j,k-1} - 4u_{jk} = -\frac{(\Delta x)^2}{k} S_{jk}
    \label{eq: C2D}
\end{equation}
The 2-dimensional space is flattened into a line, create the matrix, and obtain the following result in Figs. \ref{fig:C2D} and \ref{fig:C2D_2D}. As hyperparameters of the algorithm, $\mu=2000$ and $\tau=100$ are chosen.

The root mean square error of our tensor network from the exact inversion was $10^{-4}$ and took $4.64$ seconds to run, compared to $1.98$ ms of the PyTorch method. 

\subsection{Comparison against original HHL}
Now, the results of the TN HHL are compared against an implementation of the original HHL algorithm in Qiskit, simulated in AerSimulator with statevector evolution. The experiments consist in the resolution of 20 random sparse matrices of dimension $16\times 16$, due to the storage limitations of the statevector simulation. The matrices have a non-diagonal density of $25\%$, with all their elements generated in the interval $(-1,1)$, the same for the vectors $\vec{b}$. After their generation, the matrices are transformed to symmetric matrices and normalized by their largest eigenvalue. For resolution, the values $n_c=20$, $C=1$, $\tau=1$ and $100000$ shots are chosen for original HHL and $\tau=100$ are chosen for TN HHL, because they provide the best results for all methods. The correct result is also obtained by the PyTorch inversion function. The use of shots means that this comparison should be interpreted as a measurement-sampled HHL output reconstructed from an ideal statevector evolution; an exact ideal-HHL statevector comparison would not require sampling noise.

In the tests, the RMSE is computed after normalizing the output vectors and comparing their squared components, i.e. the probability vectors associated with the computational basis. With this convention, the mean root mean square error obtained for the TN HHL against the direct inversion is $7.5\times 10^{-3}$, against $2.6\times 10^{-2}$ for the original HHL. The mean root mean square error between the TN HHL and the original HHL is $3.1\times 10^{-2}$. Therefore, for this metric the TN HHL obtained a lower mean error against the PyTorch reference than the sampled original HHL. The TN HHL computes extremely faster than the naive statevector HHL simulation. The TN HHL needs $0.062$ seconds per problem, compared to $17.766$ seconds for the statevector. This is a $148$ times speedup. The same experiment was intended with the matrix product state backend for tensor networks, but it is extremely slower than the statevector one, and gives worse results.

\subsection{Sensitivity to hyperparameter tuning}
\begin{figure}
    \centering
    \begin{subfigure}[t]{0.48\textwidth}
        \centering
        \includegraphics[width=\textwidth]{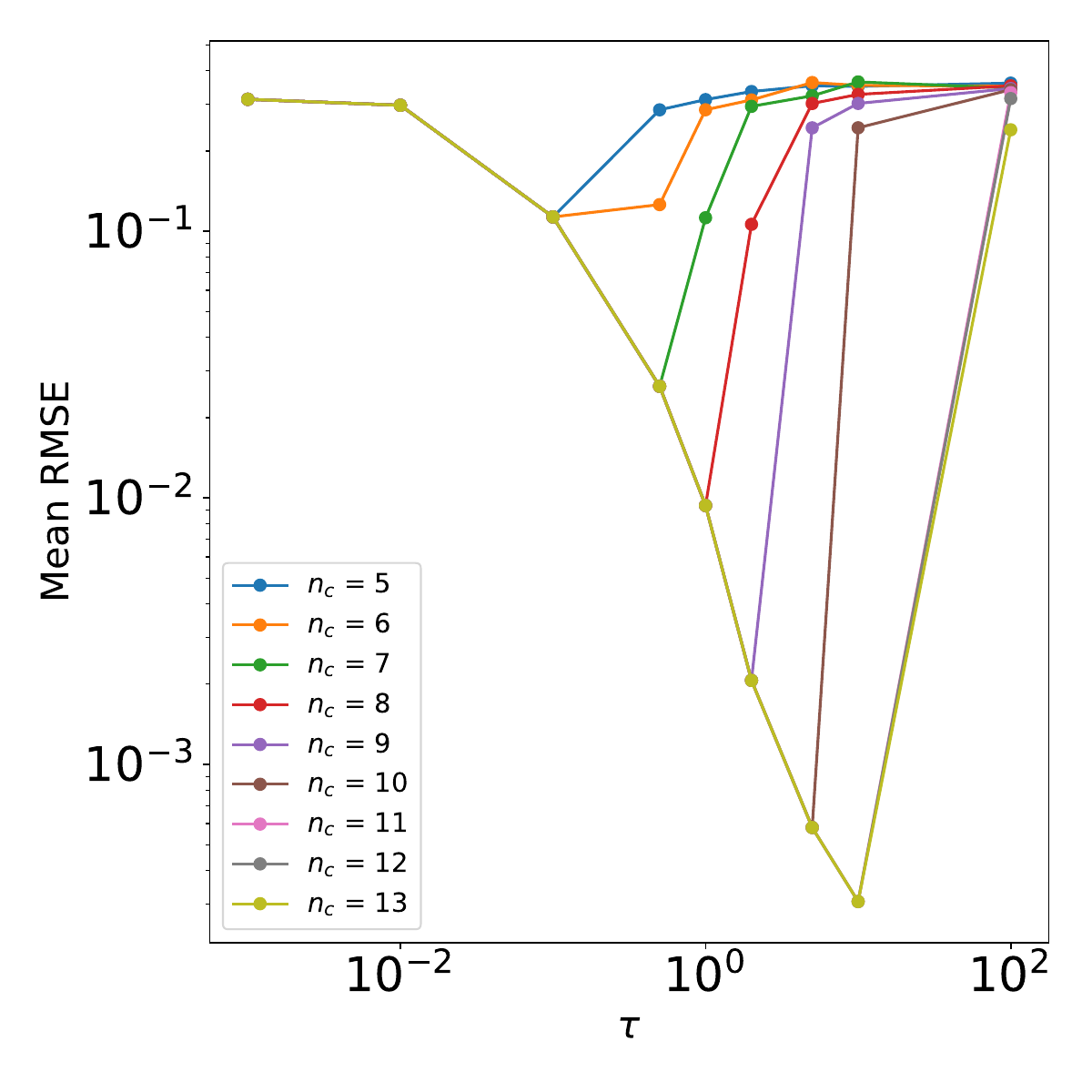}
        \caption{Mean RMSE against the $\tau$ hyperparameter for each value of $n_c$.}
        \label{fig:rmse_tau}
    \end{subfigure}
    \begin{subfigure}[t]{0.48\textwidth}
        \centering
        \includegraphics[width=\textwidth]{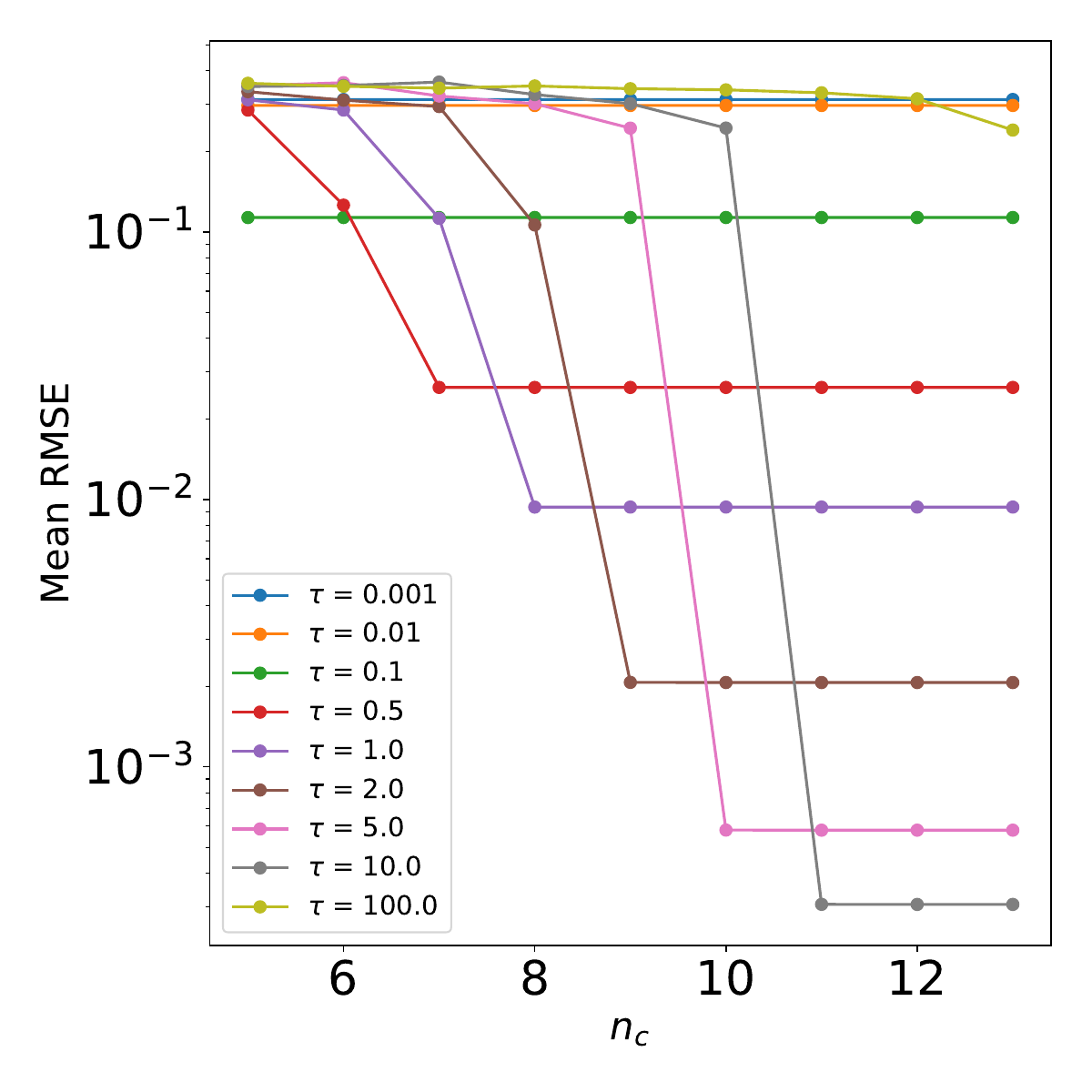}
        \caption{Mean RMSE against the $n_c$ hyperparameter for each value of $\tau$.}
        \label{fig:rmse_nanc}
    \end{subfigure}
    \caption{Comparison of Mean Root Mean Square Error (RMSE) against the hyperparameters of the HHL for random $16\times 16$ matrices. Each point is the mean of the RMSE of $20$ matrices.}
    \label{fig:rmse_comparison}
\end{figure}

Finally, this algorithm can be applied to study the sensitivity of the HHL algorithm respect to its hyperparameters $\tau$ and $n_c$. For this purpose, the same $20$ random matrices are solved with the TN HHL for different values of $\tau$ and $\mu=2^{n_c}$. Fig.~\ref{fig:rmse_comparison} shows the results of the RMSE of the prepared vector $\vec{x}$ with respect to the PyTorch inversion, after normalizing both vectors and comparing their raw components. Therefore, in this subsection the metric is applied directly to the normalized amplitudes, not to their squared components.

Fig.~\ref{fig:rmse_tau} shows that the RMSE converges in small values of $\tau$ for all values of $n_c$, to the same value. It also shows that small $n_c$ implies an increase in the RMSE, limited by normalization, but for higher values it reduces until a minimum is reached for different values of $\tau$. It seems that every $n_c$ series follows the same convergence until they reach their saturation point, which has a higher $\tau$ and lower RMSE. Also, the saturation points seem to have a similar space from the previous one (in logarithmic scale), but for $n_c>10$ they all have the same behavior. Fig.~\ref{fig:rmse_nanc} shows that the increase in the number of clock qubits always decreases or maintains the RMSE until a saturation point. Again, the optimal amount of $n_c$ is not the same for all values of $\tau$, but increases with it, which has a lower RMSE. Both results show that the HHL performance is highly dependent on the hyperparameter values, but it may be predictable with regression techniques.

\section{Conclusions}\label{sec: conclusions}
This work shows that this algorithm offers a way to efficiently simulate the noise-free tensor-network contraction associated with HHL for solving systems of linear equations, inverting matrices, and performing numerical simulations based on it. It has also been shown that its scaling is remarkably good with the size of the matrix to be inverted, while it can be realized on classical computers and accelerated with GPUs. This work also reports the existence of saturation points for the performance in the hyperparameters.

An advantage of this method is that it allows one to observe the ideal gate-error-free behavior of the HHL-inspired construction, without post-selection failures or inaccuracies in state preparation. This allows for its theoretical study prior to applying it in real noisy quantum devices or noisy simulators. At the same time, the exact inverse is recovered only under the phase-grid and no-aliasing assumptions stated in Proposition~\ref{prop:tn-exact}; outside that regime the tensor network implements the spectral filter of Proposition~\ref{prop:tn-filter}. However, the effective computational speed is remarkably low compared to methods already implemented in libraries such as PyTorch or Numpy for the resolution of the equations itself.

Future research in this area might aim to improve the overall efficiency of the method by leveraging the specific properties of the tensors used. For example, symmetry in the indices for eigenvalues in the $W[\mu]$ tensor may be applied to reduce the contraction complexity with the other tensors. This can also be particularized to tailoring the approach to tridiagonal matrices or adapting it for complex eigenvalues. Another possible line of research may be optimizing the parallelization of calculations or incorporating tensor network compression techniques to reduce runtime and storage requirements. Finally, an interesting line may be to apply this algorithm to a deeper study of the saturation points reported and the determination of a theoretical or empirical law that provides the optimal configuration of the HHL.

\bibliographystyle{unsrt}  
\bibliography{references}

\end{document}